\documentclass[twocolumn,a4paper]{article-nd2001_2}
\usepackage{nd2001_2}
\usepackage[dvips]{graphicx}
\usepackage{jnstcite}
\usepackage{nd2001-1-in-2}

\newcommand{\beq}{\begin{equation}}
\newcommand{\eeq}{\end{equation}}
\newcommand{\bcent}{\begin{center}}
\newcommand{\ecent}{\end{center}}
\newcommand{\lsim}{\mbox{\raisebox{-.3em}{$\stackrel{<}{\sim}$}}}
\newcommand{\nnb}{\nonumber}
\newcommand{\reflef}{(\ref}

\begin{document}
%%%%%  Title  %%%%%
\title{\bf Nuclear Data in Oklo and Time-Variability of Fundamental Coupling
Constants}
%%%%%  Ahuthors, affiliations  %%%%%%
\author{ Yasunori FUJII$^{1,\ast}$, Akira IWAMOTO$^2$, Tokio FUKAHORI$^2$, Toshihiko OHNUKI$^2$,\\
Masayuki NAKAGAWA$^2$, Hiroshi HIDAKA$^3$, Yasuji OURA$^4$, Peter M\"{O}LLER$^5$
\\ 
\begin{tabular}{c}
\vspace{-3.5mm}
 \\
\fontsize{9}{12}
  $^{1}$\textit{Nihon Fukushi University, Handa, Aichi, 475-0012 Japan} \\
\fontsize{9}{12}
  $^{2}$\textit{Japan Atomic Energy Research Institute (JAERI), Tokai-mura,
Naka-gun, Ibaraki, 319-1195 Japan} \\
\fontsize{9}{12}
  $^{3}$\textit{Departement of Earth and Planetary Systems Science, Hiroshima
University, Higashi-Hiroshima, Hiroshima, 739-8526 Japan} \\
\fontsize{9}{12}
  $^{4}$\textit{Department of Chemistry, Tokyo Metropolitan University,
Hachioji, Tokyo, 192-0397 Japan} \\
\fontsize{9}{12}
  $^{5}$\textit{Theoretical Division, Los Alamos National Laboratory, NM 87545, USA} \\
 \\
\end{tabular}
}

%%%%%  Abstract  %%%%%
\begin{abstract}
We re-examined  Shlyakhter's analysis of the Sm data in Oklo.
With a special care of minimizing contamination due to the inflow of
the isotope after the end of the reactor activity, we confirmed that
his result on the time-variability of the fine-structure constant,
$|\dot{\alpha}/\alpha |\lsim 10^{-17}{\rm y}^{-1}$, was basically
correct.  In addition to this upper bound, however, we obtained
another result that indicates a different value of $\alpha$ 2 billion
years ago. We add  comments on the recent result from QSO's.
%%%%%  Keywords  %%%%%
\begin{center} \begin{minipage}{145mm}
\end{minipage}\end{center}

\end{abstract}
\date{}

\maketitle

%%%%%  Footenote  %%%%%
\keyword[~$^{\ast }$]{Corresponding author, Tel. +81-569-20-0118, Fax. +81-
569-20-0127}
\keyword[~]{E-mail: fujii@handy.n-fukushi.ac.jp}

%%%%%  Text  %%%%%
More than 60 years ago, Dirac startled the science community by saying that the gravitational constant, $G$, is not a true constant, but varies as the inverse of the cosmic time $t$, and hence $\dot{G}/G \sim t^{-1}$.\cite{dir}  At the present epoch we should expect
%%%%%%%%%%%%%%%%%%%%%%%%%%%%%%
\beq
\frac{\dot{G}}{G}\sim -t^{-1}_0\sim -10^{-10}{\rm y}^{-1},
\label{dn-1}
\eeq
because the present age of the universe is $t_0 \sim 10^{10}{\rm y}$.  Unfortunately this number is so small that almost any noise would easily wipe out the real effect if there is any.

No solid evidence for the change has ever been reported.  We have
heard only about upper bounds $\lsim 10^{-11}{\rm
y}^{-1}$.\cite{vik}  Since these are already below the
prediction \reflef{dn-1}), it seems as if we have lost a motivation
for further searching for time-variability of constants.  We
emphasize, however, that we have now another ``faith,'' {\em unified
theories,} according to which having time-dependent coupling constants is a
rather natural consequence, likely to show themselves in ways quite different from what Dirac expected before.

This can be understood if we recognize that there are two different
levels of the theories.  At one of the  levels we have a fundamental
theory, like string theory as one of the best known candidates, while theories
which apply to our realistic world is at another level.  In the
former theory we are supposed to have the fundamental coupling
constants which are truly time-independent.  On the other hand, the
``effective'' coupling constants which we measure are derived basically from the coupling constants at the
deeper level.  But this process is accompanied  with  participation of
scalar fields that are expected to vary slowly as the universe
evolves.  This is the reason why we naturally expect time variation of
such non-gravitational coupling constants like the electromagnetic
charge $e$, or the fine-structure constant $\alpha = e^2/(4\pi
c\hbar)$, as well as its strong interaction analogue, $\alpha_s$.

%%%%% Table 1  %%%%%
\begin{table}[h]
{\small
\fontsize{9}{12}
 \begin{center}
   \caption{Brief summary of the past determination of
$\dot{\alpha}/\alpha$ and $\dot{\alpha}_s/\alpha_s$ in units of ${\rm y}^{-1}$.
}
   \label{tbl}
   \begin{tabular}{|l|ll|}
\hline
 Sources & $\dot{\alpha}/\alpha$ & $\dot{\alpha_s}/\alpha_s$ \\

\hline
%\vspace{-2mm} \\
Primord nucl synth\cite{hoyle} \rule[-4pt]{0pt}{14pt} &    &  $1\times 10^{-13}$    \\
 Long-lived nuclei\cite{dyson}  & $3\times 10^{-13}$  &    \\
 Stellar nucl synth\cite{davies} &    & $2\times 10^{-12}$   \\
 Oklo phenomenon\cite{sh} & $1\times 10^{-17}$ & $5\times 10^{-19}$ \\
 Time standards\cite{godone} & $3\times 10^{-13}$ & \\
 Distant QSO's\cite{webb} & $7\times 10^{-16}$ & \\

\hline
  \end{tabular}%
 \end{center}%
}%
\end{table}%

Table 1 shows a brief summary of the results of the past efforts
for $\dot{\alpha}/\alpha$ and
$\dot{\alpha_s}/\alpha_s$.\cite{hoyle}$^{-}$\cite{webb}  Most of
them are upper bounds, except for the last one from distant QSO's due
to Webb {\em et al.}, who claim to have discovered the evidence of
changing $\alpha$ for the first time.\cite{webb}  We will come
back to this issue later.  We notice; (i) the results are several
orders of magnitude below the level of $t_0\sim 10^{-10}{\rm y}^{-1}$,
(ii) ``Oklo phenomenon
``  yields much better upper bounds than others.  We are going first to discuss the second item based on our own recent re-analysis.\cite{eight}

Oklo is a name of a uranium mine in Gabon, West Africa.  This name is,
however, remembered as the
place where the evidence of  {\em natural reactors} was discovered and
verified for the first time in 1974.\cite{oklo}  By natural
reactors we mean that  self-sustained fission reactions occurred {\em
naturally} some 2 billion years ago, lasting millions of years.  Also
this was a re-discovery of the theoretical prediction due to Kuroda in
1956.\cite{kuroda}  Among several reasons for this apparently
strange phenomenon to have occurred, most crucial is that the
abundance of $^{235}{\rm U}$ was much higher ($\sim 3\%$) in 2 billion
years ago than it is $(0.7207\%)$.  Since 1974 extensive studies have
been made on the remnants of fission products in Oklo.

Then in 1976, a Russian physicist A. Shlyakhter pointed out that the
measurement of the abundance of the isotope $^{149}{\rm Sm}$ should be 
useful to determine how much nuclear reactions in the Oklo times might
have been different from what they are today.  The natural abundance
of $^{149}{\rm Sm}$ is now $13.8\%$, while the measurement in the
reactor zones showed much smaller values.  For this deficit one might  blame  the consumption due to the neutron-absorption  process:
%%%%%%%%%%%%%%%%%%%%%%
\beq
n+^{149}\hspace{-.2em}{\rm Sm} \rightarrow ^{150}\hspace{-.2em}{\rm Sm} + \gamma.\label{dn-2}
\eeq
This scenario is in fact supported  reasonably well by the use of the
data taken from today's laboratories. This implies that the process
\reflef{dn-2})  2 billion years ago should not have been very much different from what it is.

The analysis can be made even more precise by noting that the reaction
\reflef{dn-2}) is dominated by a resonance that lies as low as $E_r =
97.3 {\rm meV}$, which is more than 7 orders of magnitude smaller than
the typical mass scale $\sim 1 {\rm MeV}$ for most of the nuclear
reactions.  From this point of view, the occurrence of this resonance
must be a consequence of nearly complete cancellation between the two
effects; the repulsive Coulomb force and the attractive nuclear
force.  The former is proportional to $\alpha$, while the latter could 
 depend on $\alpha_s$ probably in a more complicated manner.   
Suppose $\alpha$ was different from the present value by $\Delta\alpha$.  No
matter how small $\Delta\alpha$ might be, it could still result in an
appreciable amount of change  $\Delta E_r$ in the residual portion
$E_r$, particularly in terms of the fractional ratio $\Delta
E_r/E_r$.  The same should be true for the cross section
$\sigma_{149}$ of the reaction \reflef{dn-2}).

This is an ideal example of an ``enhancement mechanism,'' which makes
it possible to measure something very small.  We find, as an
illustration, choosing $\Delta E_r = -10 {\rm meV}$ and $T=20^\circ
{\rm C}$ gives about $10\%$ increase of the cross section, where we
have assumed a thermal equilibrium of the neutron flux.

Unfortunately, however, his paper was too short to tell any details of the data, leaving questions on how reliable his final result was.  For this reason we decided to redo what he did previously by ourselves.\cite{eight}  We soon realized that major source of errors comes from the following situation.

Each natural reactor came eventually to an end probably in some
million years.  Even after this time, certain amount of isotopes might
have migrated from outside into the core, due to the effect of
weathering and other related phenomena.  This amount, however, has
nothing to do with the nuclear interaction we are interested in.  We
wanted to minimize this ``contamination'' as much as we could.  For
this purpose we used five samples taken from deep underground, also
with the care of geologists expertise.  In this sense we managed to
have the data of reasonably good quality.

We set up differential equations for the evolution of the system of $^{235}{\rm
U}, ^{147}\hspace{-.2em}{\rm Sm}, ^{148}\hspace{-.2em}{\rm Sm},
^{149}\hspace{-.2em}{\rm Sm}$.  By solving them we determined
$\hat{\sigma}_{149}=( 91\pm 6){\rm kb}$ for the reaction
\reflef{dn-2}), where the ``effective cross section'' is defined by
$\hat{\sigma}=\sqrt{(4/\pi)(T/T_0)}\sigma$, making it easier to handle the cross
sections obeying the $1/v$-behavior with $T_0 = 20.4^{\circ}{\rm C}$.

%%%%%%%%%%%%%%%% fig %%%%%%%%%%%%%%%%%%%
\begin{figure}[t]
\begin{center}
\includegraphics[width=8cm]{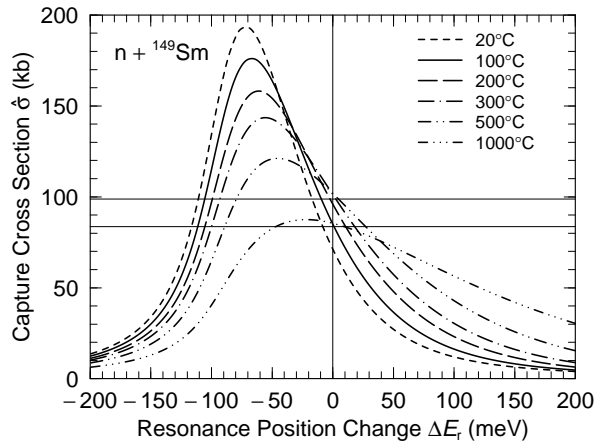}
\caption{
Thermally averaged effective cross section for ${\rm n}+ 
^{149\!\!}~{\rm Sm}\rightarrow
^{150\!\!}~{\rm Sm} + \gamma$.\cite{eight}  The horizontal two lines represent the range of the observed effective cross section $(91\pm 6)$kb. 
}
\end{center}
\label{fig1}
\end{figure}

Also we used improved methods to determine the temperature estimate;
$T= ( 200-400 )^\circ {\rm C}$.  Using these together in Fig. 1, we finally
obtain  two ranges of the allowed value of the change of the resonance energy (difference from today's value):
%%%%%%%%%%%%%%%%%%%%%%%%%%%%%%%%%%%
\[
\Delta E_{\rm r} =\left\{
\begin{array}{l}
 9\pm 11\;{\rm meV}, \hspace{1em}\nnb\\
\hspace{2em}\quad\mbox{for the right-branch range,} \nnb\\[.8em]
  -97\pm 8\;{\rm meV}, \nnb\\
\hspace{2em}\quad\mbox{for the left-branch range}. \nnb
\end{array}
\right.
\]
(The result for the right-branch range replaces Eq. (27) in reference
9, where the erroneous lower end was shown inadvertently.)  
The presence of two solutions comes from the fact that the cross
section is sharply peaked around $\Delta E_r \approx -(50-100){\rm meV}$,
depending on the temperature, giving possibly two intersections with a
horizontal line representing a value of the observed cross section.
The right-branch range, situated to the right of the peak of the cross
section, covers the zero, whereas the left-branch is away from zero.

Damour and Dyson carefully analized the Coulomb energies of
$^{149}{\rm Sm}$ and $^{150}{\rm Sm}$, finding that $\Delta\alpha$
should result in $\Delta E_r = (\Delta\alpha /\alpha){\cal M}_c$ with
the estimate ${\cal M}_c\approx -1.1{\rm MeV}$\cite{dd}.  In this way we
arrive at the two ranges of the fractional change of $\alpha$:  
%%%%%%%%%%%%%%%%%%%%%%%%%%%%%%%%%%
\[
 \frac{\Delta \alpha}{\alpha} =
\left\{
\begin{array}{l}
 -(0.8\pm 1.0)\times 10^{-8}, \nnb\\[.6em]
 (0.88\pm 0.07) \times 10^{-7}, \nnb
\end{array}
\right.
\]
and further dividing by $-2\times 10^9{\rm y}$ also at the fractional rate of change:
%%%%%%%%%%%%%%%%%%%%%%%%%%%%%%%%%%
\[
 \frac{\dot{ \alpha}}{\alpha} =
\left\{
\begin{array}{l}
 (0.4\pm 0.5)\times 10^{-17} {\rm y}^{-1},\nnb\\[.6em]
 -( 0.44\pm 0.04) \times 10^{-16} {\rm y}^{-1}.\nnb
\end{array}
\right.
\]
(The signs in Eqs. (35) and (36) of reference 9 should also be reversed.)

\lastpagecontrol{5.5cm}

The result from the right-branch gives a null result with an upper bound, as usual.  We also find that our result agrees well with Shlyakhter's previous conclusion, thus confirming his ``champion" result.  The agreement to this extent seems, however, rather accidental because it is unlikely that the data as good as ours was available in the early years of Oklo studies.

The left-branch range indicates, as it stands, the resonance energy,
and hence the value of $\alpha$, was indeed different from today's
values.  We tried to see if we could eliminate this choice by
examining other isotopes, $^{155}{\rm Gd}, ^{157}{\rm Gd}$.  The
result is less conclusive due to the stronger effect of contamination, 
nevertheless indicating a consistency with the null result of
$^{149}{\rm Sm}$.  It still seems short of coming to a final
conclusion, unless some other isotopes, probably $^{113}{\rm Cd}$, are 
included in our analysis.

Taking contamination effect more seirously seems interesting even for
$^{149}{\rm Sm}$.  If we assume the contamination of 4\%, for example,
we find $\Delta E_r = 2\pm 12\;{\rm meV}$ which replaces the previous result
with contamination ignored, hence bringing $\Delta E_r =0$ and hence
$\Delta \alpha =0$ more {\em inside}  the allowed ranges corresponding to
the 1$\sigma$-level of the cross section.

The same type of re-analysis had been made by Damour and
Dyson\cite{dd}, who obtained the range that covers basically both of
our two ranges connected without separation between them. This is due
to the larger errors in the cross section, which seems to come from
contamination that affected most of the earlier data.

We add a few comments on the recent analysis on
QSO's.\cite{webb}  They improved the analysis of past studies
based on the alkali doublets by including the effect of relativistic
corrections  to the atomic spectra, obtained from 49 QSO's, with the look-back time as old as 0.4--0.9 of the age of the universe.  They summarized the result as
%%%%%%%%%%%%%%%%%%%%%%%%%%%%%%
\[
\frac{\Delta\alpha}{\alpha} =( -0.78 \pm 0.18) \times 10^{-5},
\]
which is obviously nonzero, though some of the unknown systematic
errors are still suspected.  Comparing with this, our Oklo result
(even the nonzero result) is about 2 orders of magnitude smaller, but
the look-back time is also smaller (0.16) than the aobve mentioned
0.4--0.9.   Combining all together we
would have a non-monotonic and even oscillating time dependence, which
could be interesting from a theoretical point of view.

There are suggestions that even such traditional phenomena like big-bang nucleosynthesis and anisotropy in cosmic microwave background allow certain room for time variation of $\alpha$.\cite{ba}  Searching for time-variability of $\alpha$ will increasingly deserve further attention.
%%%%%%%%%%%%%%%%%%%%%%%%%%%%%%%%%%%%%%%%%%%%%%%%%%%%%%%%%%%%%%%%%%%%%%%%%%%%%%%

%%%%%  References  %%%%%

%%%%%%  Draw an underline  %%%%%%
%%  If there exits no right column in the final page, a line
%%  cannot be drawn in the center.
\setlength{\unitlength}{1mm}
\begin{picture}(81,10)
 \thicklines
\put(-47.5,8){\line(1,0){81}}
%\put(0,8){\line(1,0){81}}
\end{picture}

\end{document}